\documentclass[final]{svjour2}
\usepackage{graphicx}
\usepackage{rotating}
\usepackage{amssymb}
\usepackage{mathptmx}
\usepackage[numbers]{natbib}
\usepackage{color}
\makeatletter
\journalname{Journal of Low Temperature Physics}

\bibpunct{}{}{,}{s}{}{,}

\begin{document}

\newcommand{\hdblarrow}{H\makebox[0.9ex][l]{$\downdownarrows$}-}
\title{Superfluidity of flexible chains of polar molecules}

\author{B. Capogrosso-Sansone$^1$ \and A. Kuklov$^2$}

\institute{1:Institute for Theoretical Atomic, Molecular and Optical Physics,
Harvard-Smithsonian Center of Astrophysics,\\ Cambridge, MA 02138, USA\\
\email{bcapogrosso@cfa.harvard.edu}\\
2:Department of Engineering Science and Physics,
CSI, CUNY, \\Staten Island, NY 10314, USA\\
\email{Anatoly.Kuklov@csi.cuny.edu}}
\date{06.05.2011}

\maketitle


\begin{abstract}
We study properties of quantum chains in a gas of polar bosonic molecules confined in
a stack of $N$ identical one- and two- dimensional optical lattice layers, with molecular dipole moments aligned perpendicularly to the layers. 
Quantum Monte Carlo simulations of a  single chain (formed by a single molecule on each layer) reveal its quantum roughening transition. The case of finite in-layer density of molecules is studied within the framework of the J-current model approximation, and it is found that  $N$-independent molecular superfluid phase can undergo a quantum phase transition to a rough chain superfluid. A theorem is proven that  no superfluidity of chains with length shorter than $N$ is possible. The  scheme for detecting chain formation is proposed.  

PACS numbers:
\end{abstract}

\section{Introduction}
Ultracold polar
molecules, interacting via dipolar forces which can be tuned and shaped by external static and microwave electric
fields~\cite{Micheli}, represent a
fertile ground for the study of exotic phases of matter. In recent experiments, high phase space
density of both polar and non-polar molecules in stable ground states have been realized~\cite{Jila1, Nagerl1}. The JILA group has also reported on observations of
dipolar collisions and an evidence of spatial anisotropy of dipolar interaction in a 
sample of KRb molecules~\cite{Jila2}.
The Innsbruck group has produced a high density sample of $\rm{Cs_2}$ molecules 
 in an optical lattice~\cite{Nagerl1}. The procedure used in~\cite{Nagerl1} 
can be generalized to heteronuclear (polar) molecules~\cite{Nagerl2}. 

There exist many predictions of exotic phases formed by polar molecules -- self-assembled dipolar crystals, dipolar Bose-Einstein condensates, supersolids ~\cite{Buchler-Baranov,BCS}. Another interesting proposal concerns polar molecules in optical lattice bilayers with no inter-layer tunneling and induced dipole moments perpendicular to the planes~\cite{Trefzger}. Such system is expected to display paired-supersolidity, that is, supersolid of composite objects formed by two molecules, each belonging to different layer. Given successful experimental advances, these phases are very likely to be within reach in near future.   Here we consider a multilayered system and study the formation of quantum {\it flexible} chains of dipolar molecules.
\\ \indent
Physical properties of extended quantum objects are of great interest to many areas of physics. Quantum strings described by first-quantized Nambu action are at the base of modern high energy physics~\cite{strings_HE}. In condensed matter, significant examples of non-point-like quantum objects include dynamical stripes in high-Tc superconductors~\cite{stripes}, quantum vortices in superfluids and superconductors, dislocations in quantum crystals, and chains of polar molecules~\cite{Wang, layers}. Self-assembly of {\it classical} dipolar particles, especially in the context of colloids, e.g., ferrofluids, has been investigated quite extensively (see Ref.~\cite{Lozovik} and references therein). Recently, {\it quantum} chains of ultracold bosonic polar molecules have been addressed. Thermal and quantum fluctuations of chains of polar molecules forming solid structures in highly anisotropic traps were studied by Monte Carlo ~\cite{Boronat}. In Ref. ~\cite{Wang}, the possibility of Bose-condensation of {\it stiff} non-interacting chains of polar bosonic molecules formed in a stack of pancakes with zero inter-layer hopping has been proposed. In this simplified approach the intra-chain dynamics as well as molecular exchanges between different chains have been ignored. The fermionic case was studied in Ref. \cite{layers}. 
\\ \indent In the present work we focus on superfluid properties of flexible and interacting chains of polar molecules  in optical lattices.  The single chain as well as the chain "soup", with allowed molecular exchanges between different chains, have  been studied. Our results are based on Monte Carlo simulations by the Worm algorithm~\cite{WA} and can be summarized as follows: i)  A single chain, formed by one molecule per layer, undergoes quantum roughening transition driven by the presence of the optical lattice potential within the layers; ii) An ensemble of flexible chains can exhibit superfluidity characterized by off-diagonal long-range order (ODLRO) in the $N$-body density matrix only ; iii) No superfluidity of chains shorter than the number of layers $N$ can exist, provided that all $N$ layers are equivalent to each other; iv) There exist a quantum phase transition from the chain superfluid (CSF) to $N$ independent molecular superfluids (N-SF) phase. The latter is induced by molecular exchanges between different chains. For two dimensional (2d) layers this transition is strongly  first-order, while for one-dimensional (1d) layers it is continuous.
\section{The model}
We consider a stack of  $N$ parallel to each other optical lattice layers -- either 2d planes or 1d "cigars", populated by bosonic polar molecules, 
with dipole moments aligned by an external static electric field perpendicularly to the layers (along $z$-axis) . At low enough temperature the system is well described by the Hamiltonian:
\begin{equation}
H= - J \sum_{\langle ij\rangle, \alpha} a^\dagger_{i\alpha}a_{j\alpha} + \frac{1}{2}\sum_{i\alpha; j\gamma}V_{i\alpha;j\gamma}n_{i\alpha}n_{j\gamma},  
\label{Habi}
\end{equation}
where $a^\dagger_{i\alpha},\, a_{j\alpha}$ are creation-annihilation operators of a boson at site $i$ and on layer $\alpha=1,2,...,N$ (Latin indexes 
label sites within a layer, Greek indexes label layers); $J$ stands for tunneling amplitude between neighboring sites within each layer;  
$n_{i\alpha}=a^\dagger_{i\alpha} a_{i\alpha}$ denotes the onsite density operator obeying the hard-core constraint; $V_{i\alpha;j\gamma}$ 
describes the dipolar interaction matrix elements between sites $(i\alpha)$ and $ (j\gamma)$. This interaction is characterized
by a typical strength $V_d=d_z^2/b^3_z$, where $d_z$ stands for the induced dipole moment along the $z$ direction, and $b_z$ denotes the distance between two nearest layers. The anisotropic dipole-dipole interaction is repulsive within layers and mainly attractive between layers.
Simulations have been conducted in the situation of zero inter-layer tunneling.

\subsection{J-current version}
While we use model~(\ref{Habi}) for studying a single chain formed by one molecule per layer, the many body case, i.e., finite density $n$ of molecules on each layer, has been analyzed within the approach which is a multi-component version of the J-current model \cite{Jcurrent}:
\begin{equation}
H_J = \sum_{b\alpha}\left[ \frac{K_1\vec{J}^{\,2}_{b\alpha}}{2} +\frac{K_2(\nabla_\alpha\vec{J}_{b\alpha})^2}{2} - \mu J^{(d+1)}_{b\alpha}\right], %
\label{HJ}
\end{equation} 
where $\nabla_\alpha \vec{J}_{b\alpha} \equiv \vec{J}_{b (\alpha +1)} - \vec{J}_{b\alpha}$;
$\vec{J}_{b\alpha}$ stands for an integer vector along bond $b$ between two neighboring sites of the
$D=d+1$ space-time lattice in the $\alpha$-th layer (each layer is characterized by dimensions $L_\nu =L,\, \nu =1,...,d,\,$ and $L_{d+1}=\beta$, with $\beta$ the inverse temperature); $\vec{J}_{b\alpha}$ obeys Kirchhoff's conservation law $ \vec{\nabla}_b \vec{J}_{b\alpha}=0$ (absence of interlayer
tunneling implies no currents perpendicular to the layers);
$1/K_1 \sim J$ denotes bare superfluid stiffness of molecules in each layer; $K_2 \sim V_d/J$
describes attractive coupling between neighboring layers; the  chemical potential $\mu$ is the same for all layers. 

Model~(\ref{HJ}) represents a dual formulation of the Bose-Hubbard model, with $\vec{J}_{b\alpha}$ carrying a meaning of an element of particle world-line in discretized time. Such approach provides qualitative insights on phases and phase transitions occurring in the original quantum Hamiltonian. We are utilizing it as a first step toward full scale quantum {\it ab initio} simulations.  In this work we consider the simplest version -- the interactions are cut off to nearest-neighboring sites on nearest neighboring layers and all intra-layer terms are ignored.  For $K_2=0$ and at $\mu=0$ this model exhibits a phase transition from an insulating state to a superfluid one, the latter being characterized by macroscopically large closed loops formed by the J-currents \cite{Jcurrent}. 

In the present context of $N$-layered geometry there are $N$ independent superfluids ( that is, N-SF phase) for $K_2=0$. In the opposite limit , i.e. at large enough $K_2$, it becomes energetically favorable for world-lines in neighboring layers to lock in together, so that $\nabla_\alpha \vec{J}_{b\alpha} \approx 0$.  It will be shown that such locking indicates formation of chains of length $N$. In other words, there is no superfluid phase of chains of length $1<M<N$. 

In what follows we will be discussing the hard-core version of the J-current model, that is, the condition $|\vec{J}_{b\alpha}| =0,1 $ is imposed. This implies that the filling factor per each layer is $0<n<1$ and there is always $N$-SF phase at $K_2=0, T=0$. The situation can change in the presence of significant interaction between neighboring sites within the layers induced by the dipole-dipole coupling. Such interaction can induce solid phases competing with superfluidity. We will consider such possibility in the future. 

\section{Quantum roughening of single chain}
Let us start by discussing the case of a single chain formed by one dipole 
per layer. We have performed quantum Monte Carlo simulations of the model~(\ref{Habi}) for a stack of both 1d cigars and 2d layers with linear size $L$. Hard wall conditions have been imposed at the space edges $\pm L/2$ of each layer, with one of the dipoles pinned in the middle of the layer it belongs to. 

Since a two-body bound state in bilayers is to exist for arbitrarily small dipolar interaction strength~\cite{Schain}, the chain is to be formed at any finite $V_d/J$, provided $L\to \infty$ and $\beta \to \infty$. Practically, simulations are conducted at finite $L$ and $\beta$, and therefore, there is a threshold interaction $V_d/J$ for chain formation which is strongly dependent on $L$. It will be discussed below that such threshold is independent on $L$ when the density of molecules is finite.   
\\ \indent
We characterize chain roughness by the square of the gyration radius $R_g$:
\begin{equation}
R^2_g= \frac{1}{N} 
\sum_{\alpha=1,2,...,N}\langle \vec{r}^{\,2}_\alpha \rangle \; ,
\label{Gyr}
\end{equation}
where $\vec{r}_\alpha$ stands for the spatial position of the dipole in the $\alpha$-th layer and $\langle ...\rangle$  implies full quantum-thermal averaging. When the interaction is turned off, the positions of particles are completely uncorrelated and $R^2_g=R^2_0\equiv \sum_{i} \vec{x}^2_i /L^d \approx d L^2/12$, where $\vec{x}_i$ are the lattice sites. There exist two chain phases -- {\it rough} and {\it smooth} --  which are separated by a quantum phase transition. 
In the limit of strong interactions the chain is smooth (stiff), so that the relative displacements of particles are of the order of the lattice spacing $a$, i.e., $R^2_g=R^2_{\rm stiff} \approx a^2$, and they concentrate around the pinned dipole. As $V_d/J$ decreases, the chain enters the rough state where $R^2_g$ diverges logarithmically with $N$ (see below), but it remains significantly smaller than $R^2_0$. The phase transition is characterized by a change in the excitation spectrum -- from gapped in the stiff regime to sound-like modes propagating along the chain length (the $z$-direction) in the rough phase.

\subsection{Elastic string model}
\indent In order to gain  a qualitative understanding, we model the chain as a gaussian elastic string subjected to the lattice potential. 
 Let us consider the case of one dimensional optical lattices along the $x$-direction, that is, the displacement of the string is along $x$, while the sound propagates along $z$. The quantum roughening is controlled by the weight $\exp(-S)$, where the action in units of $\hbar$ reads as:
\begin{equation}
S=\frac{1}{\hbar} \int_0^{\beta}d\tau\int_0^{L_z}dz\left [\frac{\rho}{2}\left(\frac{\partial x(z,\tau)}{\partial \tau}\right)^2+\frac{\rho}{2} V_s^2\left(\frac{\partial x(z,\tau)}{\partial z}\right)^2-U {\rm cos}\left(\frac{2\pi}{a}x\right)\right ]
\label{S_string}
\end{equation}
where $\tau$ is the imaginary time; $x(z,\tau)$ denotes the displacement of the dipole belonging the layer $z$ at imaginary time $\tau$; $\rho=m/b_z$ is the linear mass density ($m=\hbar^2/2Ja^2$ is the effective mass in tight binding approximation); $U$ is the depth of the optical lattice potential; $V_s$ stands for the speed of sound along the string. In the strongly interacting regime $V_s$ can be expressed in terms of the dipolar interaction strength: $V_s^2\approx V_d/(\rho b_z)$. 

Rescaling $x=a \phi/(2\pi)$, $ V_s \tau \to \tau$ (and $\beta \to \beta V_s$), we arrive at the canonical form of the sine-Gordon model~\cite{SG} action:
\begin{equation}
S_{\rm{SG}}=\int_0^{\beta}d\tau\int_{-\infty}^{\infty}dz\left \{ \frac{1}{8\pi K}\left [\left(\frac{\partial \phi(z,\tau)}{\partial \tau}\right)^2+\left(\frac{\partial \phi(z,\tau)}{\partial z}\right)^2\right ]-\tilde{U} {\rm cos} \phi\right \}
\label{S_SG}
\end{equation} 
where $K=\frac{\pi \hbar}{\rho a^2 V_s}  $ is the Luttinger liquid parameter and $\tilde{U}= U/\hbar V_s$. The roughening transition of Berezinskii-Kosterlitz-Thouless (BKT) type occurs at $K=K_c=2$, so that for $K>K_c$
the $\cos$-term can be essentially ignored. In terms of the microscopic parameters of the original Hamiltonian $J,V_d, a, b_z$, we estimate $K_c \approx \frac{b_z}{a} \sqrt{\frac{J}{V_d}}\approx 1$, where the numerical coefficient of order of unity has been ignored. 
\\ \indent 
Starting from (\ref{S_SG}) one can evaluate the gyration radius, Eq.(\ref{Gyr}), in the rough regime, where the $\cos$-term can be neglected.
Then, the action becomes gaussian and $R^2_g=\int d\tau \int dz \langle x^2(z,\tau)\rangle/(b_zN\beta)= \int d\omega \int d q\langle \tilde{x}^2(q,\omega)\rangle \sim \int d\omega \int d q /(\omega^2 +q^2)\sim \int^\infty_{q_0} dq/q$, where
$ \tilde{x}(q,\omega)$ is the Fourier transform of $x(z,\tau)$ and the summation over Matsubara frequencies $\omega$ is replaced by integration in the limit T=0.
The resulting divergence is to be cut off at momentum $q\sim q_0 \approx 1/b_zN$. This leads to the log-type divergence
$R^2_{g} =R^2_q \approx \frac{\hbar  b_z}{2\pi m V_s} \ln N \propto \ln N$. 

At high $T$, while the chain still exists, the string action (\ref{S_SG}) becomes classical: $S_{cl}=\beta \int dz \rho V_s^2(\partial x /\partial z)^2/2$. Accordingly, the gyration radius (\ref{Gyr}) can be evaluated as a simple thermal average $R^2_g=R^2_{cl}= \int\, d z  \langle x^2(z) \rangle/(Nb_z)$. In Fourier space, $R^2_{cl}= \frac{Tb_z}{2\pi mV^2_s} \int^\infty_{q_0} \frac{dq}{q^2} \approx T b_z^2 N/(m V^2_s) \propto N$.

\subsection{Quantum Monte Carlo results}
\begin{figure}
\begin{center}
\includegraphics[%
  width=0.65\linewidth,
  keepaspectratio]{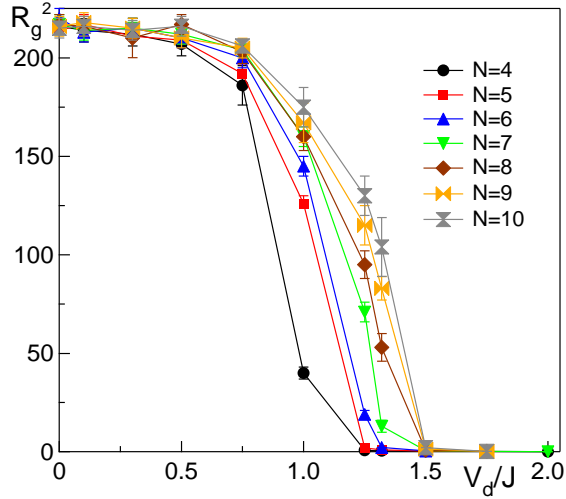}
\end{center}
\caption{(Color online)
 $R^2_g$ as a function of $V_d/J$ in the two-dimensional case. The dipole potential has been cut off to nearest neighboring sites.
The system sizes are $L=36$ and $\beta=36$, $b_z=a$. Roughening occurs at $(V_d/J)_c =V_c\approx 1.1 \pm 0.1$. For $V_d/J \to 0$, $R_g$ approaches the independent layers limit $R^2_0\approx 216$ and, for $V_d/J >V_c$, it becomes $\approx 1$.} 
\label{2d}
\end{figure}

\indent
We have performed quantum Monte Carlo simulations of the model (\ref{Habi}) in the $T=0$ limit, and have found that the numerical results for $R^2_g$ are consistent with the log- type divergence with respect to $N$ in both 1d and 2d cases. 
Fig.\ref{2d} shows the occurrence of the transition at $(V_d/J)_c \approx 1.1 \pm 0.1$ for a system of 2d layers of linear size $L=36$ (we have used a cutoff to the nearest neighboring sites for $b_z=a$). At low interaction strength, $R^2_g$ approaches the expected value for non-interacting particles $R^2_0=216$ (at L=36). For large $V_d/J$, $R_g$ is independent on the number of layers and approaches $a^2$, as one would expect for the stiff chain. We have also studied the roughening behavior as a function of the cutoff, and observed that it is weakly dependent on it. In other words, the dipole-dipole interaction can essentially be replaced by the nearest-layer interaction for a single chain.
\section{Multi-chain condensate}
We now turn to the discussion of the case of finite density $n$ of dipoles per layer. The existence of a regime where a single chain is rough immediately raises the possibility of  superfluidity of an ensemble of chains -- CSF. It is important to note that, in contrast to the single chain case where there is no threshold for $V_d/J$ to induce binding of dipoles into a chain, molecular exchanges between different chains at finite $n$ destroy CSF in favor of the N-SF phase for  $V_d/J$ smaller than some finite value. 

\subsection{Chain superfluidity order parameter}

In general, ODLRO of chains of length M can be defined via the $M$-body density matrix $D_M$, in turn defined in terms of $M$ bosonic operators $\psi_\alpha(\vec{x}_\alpha)$ placed  on $M$ adjacent layers:
\begin{equation}
D_M(\vec{x}_1, ..., \vec{x}_M; \vec{x})=\langle \Phi^\dagger(\vec{x}_1,...,\vec{x}_M) \Phi(\vec{x},\vec{x},...,\vec{x})\rangle \; , 
\label{D_M}
\end{equation}
where $\Phi(\vec{x}_1,...,\vec{x}_M) =  \psi_1(\vec{x}_1)\psi_2(\vec{x}_2)\cdot ...\psi_M(\vec{x}_M)$.
If $D_M$ is both, short ranged with respect to relative distances of the first set of $M$ coordinates $ \vec{x}_1, ..., \vec{x}_M$ and long ranged with respect to the distance between the center of mass $R_{cm} = [\vec{x}_1 + ... + \vec{x}_M]/M$ and $ \vec{x}$, and all others $D_P$ are short ranged, then $\Phi(\vec{x}_1,...,\vec{x}_M)$ describes a condensate of chains of length $M$. Accordingly, the N-SF phase is characterized by ODLRO in $D_1(\vec{x}_\alpha ; \vec{x}'_\alpha)$ ($\alpha=1,2,...,N$), with all $D_M$ ($M\leq N$)trivially long-ranged as they can be factorized into products of $D_1$ (note though, that $D_M$ will not be short ranged with respect to relative distances of the first set of $M$ coordinates).

 One may wonder whether it is possible to realize a condensate of chains of length $M<N$. This can occur quite trivially, if layers are not equivalent to each other. As an example, consider the situation where some of the layers, say the upper half, are separated by a larger $b_z$, implying negligible binding interaction. Thus, while retaining its integrity in the lower half, the chains will partially decay into $N/2$ single-particle superfluids on the upper half. We do not address such situations and consider all layers to be equivalent to each other, with the same average filling $n$ per site. We prove that under these conditions one can only have condensation of N long chains.

Formation of ODLRO gives rise to classical fields corresponding to broken U(1) symmetries. The N-SF phase is characterized by $[{\rm U}(1)]^N$ broken symmetries, that is, by $N$ (quasi-) condensed phases $\varphi_\alpha$ of $N$ U(1)-order parameters. Condensate of M-long chains is characterized by the condensed phase $\Phi_M=\sum_{\alpha=1,2,...,M<N} \varphi_\alpha$.  For identical layers, though, any choice of $M$ adjacent layers where the condensed chains will form is equivalent (we don't consider a possibility of exotic long-range interactions which can induce pairing between non-adjacent layers). Therefore, assuming spatial symmetry between the layers, 
one concludes that there is at least $N-1$ constraints on the phases, which imply that all $N$ phases are condensed, i.e., the N-SF. Alternatively, the spatial symmetry between the layers could have been broken spontaneously so that the chains, say, $M=N/2$ are formed on the upper $N/2$ layers and on the lower $N/2$ layers, with the plane of zeros separating the two $M=N/2$-condensates.
 This option, however, is excluded by Feynman's argument forbidding the ground state of bosons in time-reversal symmetric Hamiltonian to have zeros \cite{Feynman}. Thus, the only option left is $M=N$, which implies the condensation of the field $\Phi= \psi_1(\vec{x}_1)\psi_2(\vec{x}_2)...\psi_N(\vec{x}_N)$.

\subsection{Monte Carlo results}
\indent 
We have performed Monte Carlo simulations of the model (\ref{HJ}) within a multi-worm version of the Worm algorithm \cite{WA} applied to hard-core particles, i.e. $|J^{(\nu)}_{b\alpha}|=0,1$. The calculated quantities are: the superfluid stiffnesses $\tilde{\rho}_\nu,\, \nu=1,..., d$, and the gyration radius $G(N)$. $\tilde{\rho}_\nu$ are determined by fluctuations of total J-currents $I^{(\nu)}_\alpha =\sum_b J^{(\nu)}_{b\alpha}$  in a given spatial direction $\nu$ of the space-time lattice \cite{Ceperley}. 
In the independent molecular superfluid phase for $V_d=0$, there are no correlations between currents, that is, $\langle I^{(\nu)}_\alpha I^{(\nu)}_{\alpha'}\rangle \sim \delta_{\alpha,\alpha'}$.
As $V_d$ is increased, correlations appear indicating a non-viscous inter-layer drag effect. Upon further increasing $V_d$ beyond some critical value, the rough chain superfluid phase occurs, and, in the thermodynamic limit, the inter-layer correlator becomes independent of the layer index.  
This implies $\langle [I^{(\nu)}_\alpha - I^{(\nu)}_{\alpha'}]^2\rangle/(L^d \beta) \to 0$, 
that is, full locking of superflows 
in each layer into a single flow, so that no super-counter-flows in different layers occurs.
\\ \indent 
It is convenient to introduce the Fourier transform of the total J-currents with respect to the layer index:  
\begin{equation}
\tilde{\vec{I}}(q_z)=\frac{1}{\sqrt{N}}\sum_{\alpha=1,2,...,N} \exp(iq_z \alpha)\vec{I}_\alpha\; ,
\label{J_FT}
\end{equation}
where we have used a vector notations for the currents and their transforms; and $q_z= 2\pi n_z/N,\,\, n_z=0,1, 2, ...,N-1$. 
We define the superfluid stiffnesses as
\begin{equation}
  \rho^{(\nu)}(q_z) =\frac{1}{L^d\beta}\langle  \tilde{I}^{(\nu)}(q_z) \tilde{I}^{(\nu)}(-q_z)\rangle_{D_N},          
\label{Fur}
\end{equation}
where $\langle ...\rangle_{D_N}$ implies averaging over $D_N$. For all practical purposes Eq.~(\ref{Fur})  is equivalent to the well known definition in terms of windings \cite{Ceperley}.

Due to full locking of superflows in each layer into a single flow, the rough chain superfluid phase is characterized by $ \rho^{(\nu)}(q_z) \sim \delta_{q_z,0}$. In the N-SF phase, instead, full locking of superflows is lost but correlations between currents $ \rho^{(\nu)}(q_z) $ persist, resulting in a smooth dependence of $ \rho^{(\nu)}(q_z)$ on $q_z$, with a non-zero minimum at $q_z=\pi$. The quantum phase transition occurs 
when $\tilde{\rho}^{(\nu)}(\pi)\to 0$  and $\tilde{\rho}^{(\nu)}(0) \to const \neq 0$, with increasing $L$ and $\beta$. In other words, in the thermodynamic limit the N-SF
phase coherence lengths $\xi$ along the layers diverge for all harmonics $q_z$. In contrast, in the CSF phase $\xi $ becomes finite for all but $q=0$ harmonics.   
 \begin{figure}
\begin{center}
\includegraphics[%
  width=0.75\linewidth,
  keepaspectratio]{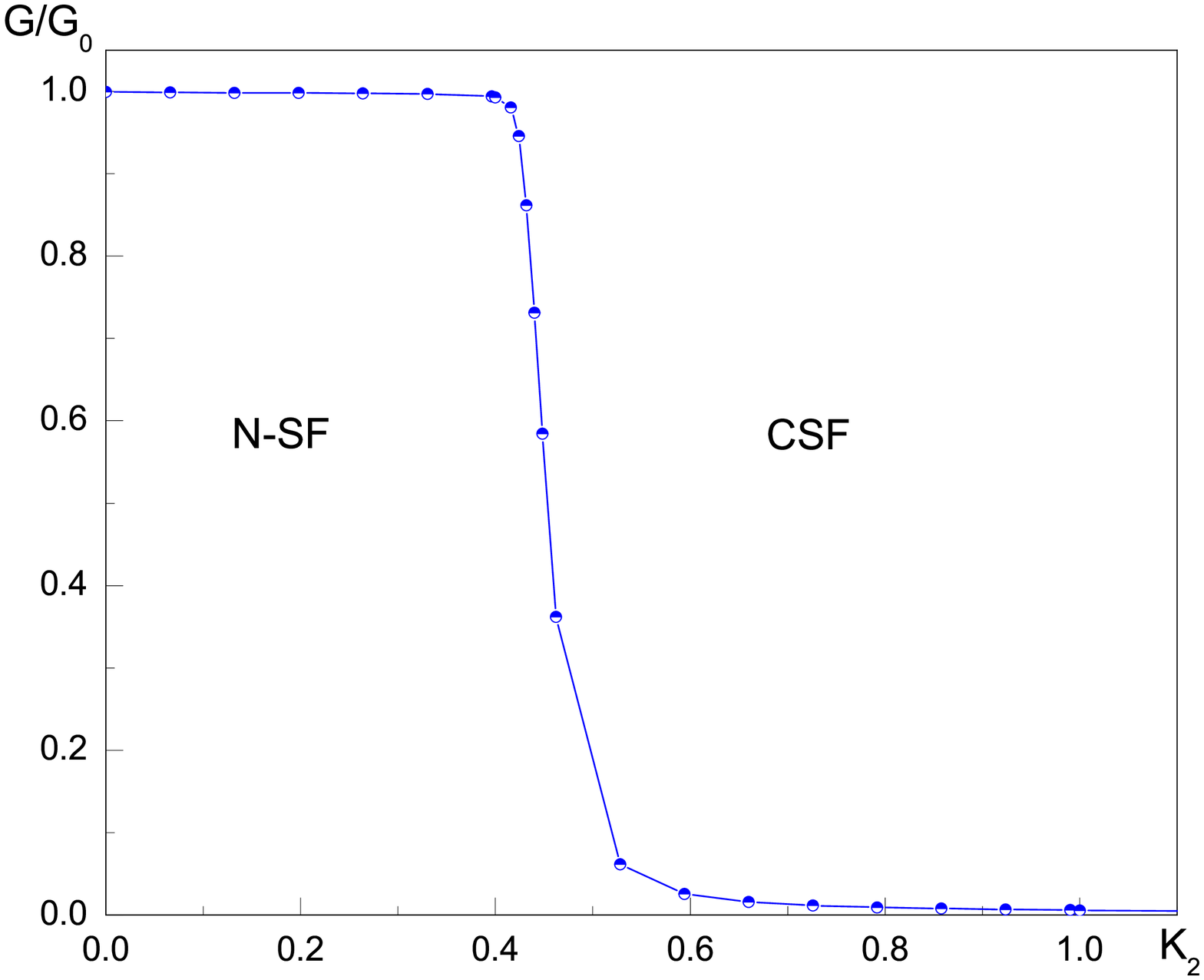}
\end{center}
\caption{(Color online)
Gyration radius $G$ vs $K_2$ normalized by its value $G_0\approx 1270$ at $K_2=0$. The parameters are $d=1,\,n=0.5,\, N=10,\, \beta=130, \, L=130$.
The transition N-SF to CSF occurs at $K_2\approx 0.42$.} 
\label{GG0}
\end{figure}
\\ \indent
The second observable we have used to characterize the system is the square of the gyration radius $G(N)$, defined as follows for the multi-chain ensemble:   
\begin{equation}
G(N)= \frac{1}{N^2} \sum_{\alpha, \alpha'}\langle \left[ \vec{x}_\alpha - \vec{x}_{\alpha'}\right]^2\rangle_{D_N}\; .
\label{Gyrn}
\end{equation}
$G(N)$  describes the lateral spread of the chain $N$-body wave-function. Such definition is a natural generalization of $R_g^2$, Eq.(\ref{Gyr}), for the single chain case. 
In the NSF phase $G(N)=G_0= (1-1/N) R^2_{0}\approx d L^2/12$ in full analogy to $R^2_0$. Fig.~\ref{GG0} demonstrates that, at a given N (here N=10), the behavior of $G(N)$ as a function of the interaction strength $K_2$ is similar to
the case of the single chain shown in Fig.~\ref{2d}. 
(Nonetheless, as we shall discuss below, unlike the single chain case, in the CSF phase chains remain rough even upon increasing the dipolar interaction strength $K_2$.)
Beyond the N-SF to CSF transition point $(K_2)_c$, the gyration radius  $G(N)$ demonstrates the quantum rough chain behavior, i.e. a log divergence with the number of layers N (see below). In order to numerically observe such features the system size has to be increased so that $ L > \sqrt{G(N)} \sim \sqrt{\ln N}$, and $L >\xi$ for $q_z\neq 0$.

\begin{figure}
\begin{center}
\includegraphics[%
  width=0.95\linewidth,
  keepaspectratio]{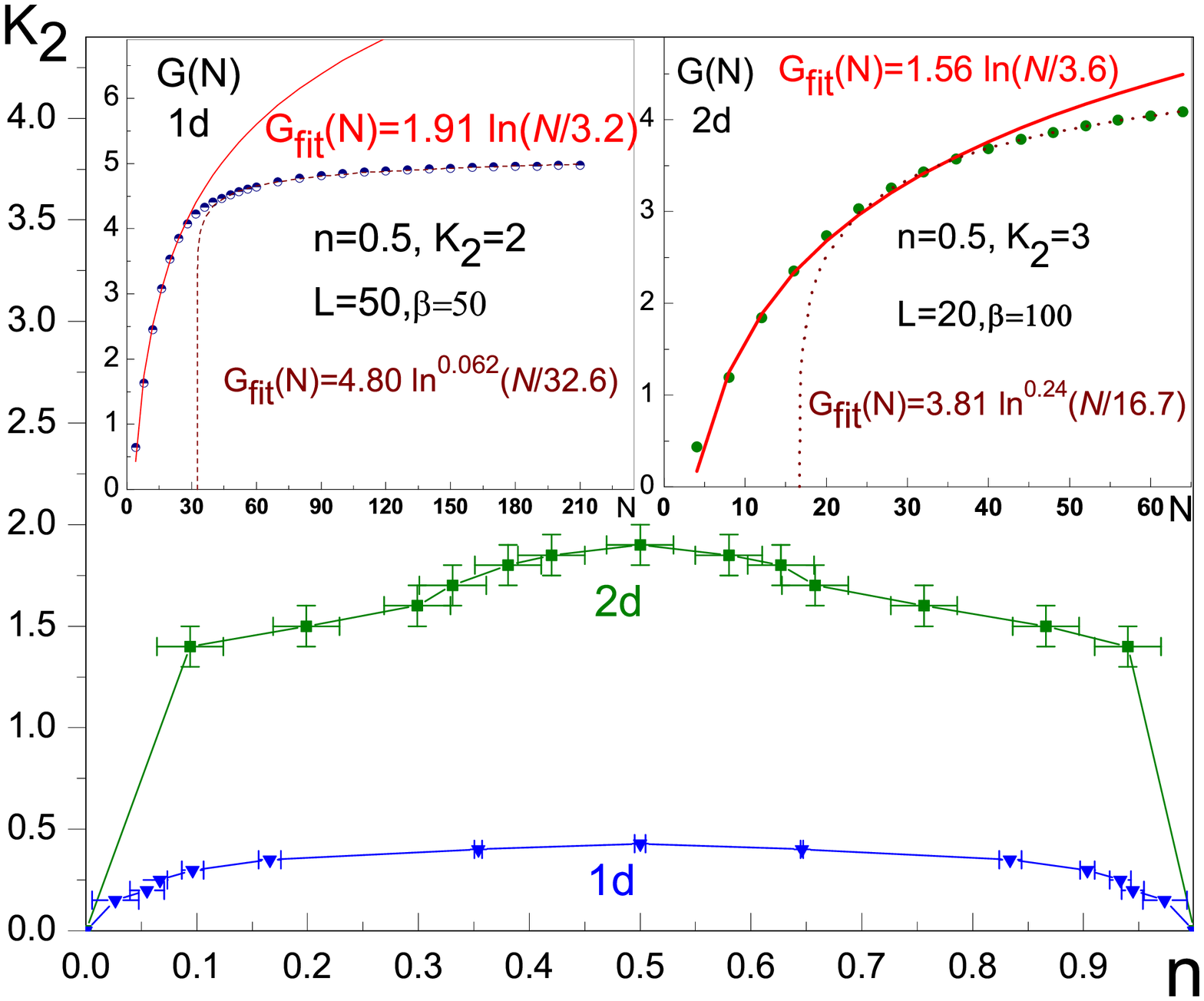}
\end{center}
\caption{(Color online)
Phase diagram within model (\ref{HJ}) in the $K_2$ vs $n$ plane in 2d ($L=24,\,N=16,\, \beta=24$, squares), and in 1d ($L=130,\, \beta=130, \, N=10-40$, triangles). N-SF (CSF) lies below (above) the transition line. 
Insets: gyration radii $G(N)$, Eq.(\ref{Gyrn}), (error bars are smaller than
symbols) and their fits (lines) in 1d  ( $R^2_{0}\approx 208$) and 2d ( $R^2_{0}\approx 67$).} 
\label{PD}
\end{figure}

\begin{figure}
\begin{center}
\includegraphics[%
  width=0.95\linewidth,
  keepaspectratio]{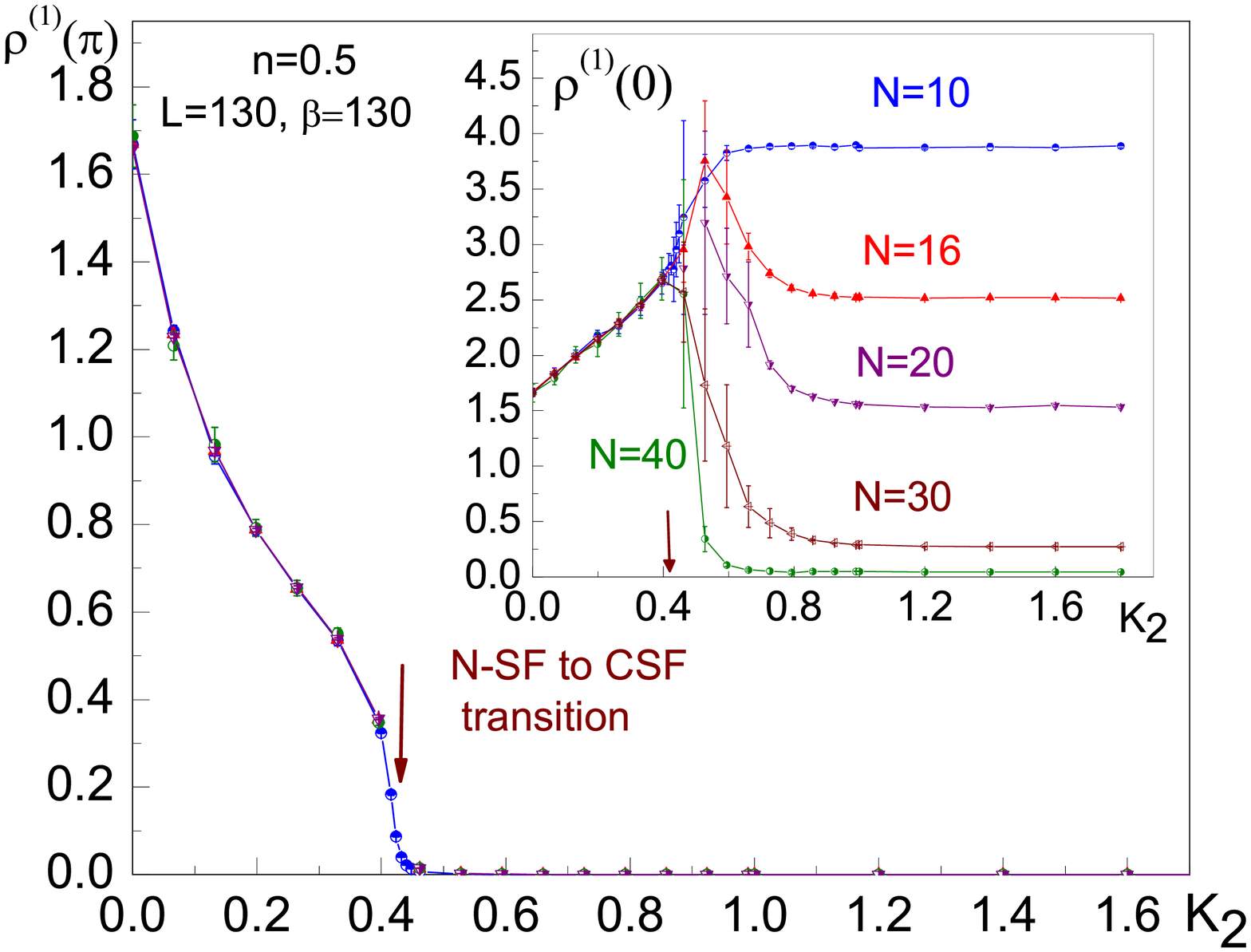}
\end{center}
\caption{(Color online)
Spatial counterflow superfluid stiffness $\rho^{(1)}(q_z=\pi)$ vs $K_2$ in the 1d-layers case ($L=130,\, \beta=130, \, N=10,20$). The vertical arrow indicates N-SF to CSF transition. Inset: the chain spatial superfluid stiffness $\rho^{(1)}(q_z=0)$.  
} \label{rho}
\end{figure}

Our numerical results are summarized in Fig.~\ref{PD}. We have chosen the bare superfluid stiffness $1/K_1=5.0$, and considered the case of hard core bosons. The main plot of Fig.~\ref{PD} shows the transition line for the 1d (triangles) and 2d (squares) systems of linear sizes $L= 130$, $\beta=130$, $N= 10-40$, and $L= 24$, $\beta=24$, $N= 16$, respectively. Upon increasing the strength of the interlayer attraction (terms $\propto K_2$ in Eq.(\ref{HJ})), chain superfluidity emerges, with a transition point which depends on the interlayer molecular density $n$. The transition lines have been determined according to the criterion described above, i.e., $\tilde{\rho}^{(\nu)}(\pi)\to 0$  and $\tilde{\rho}^{(\nu)}(0) \to const \neq 0$ as demonstrated in Fig.~\ref{rho}. 
\begin{figure}
\begin{center}
\includegraphics[%
  width=0.95\linewidth,
  keepaspectratio]{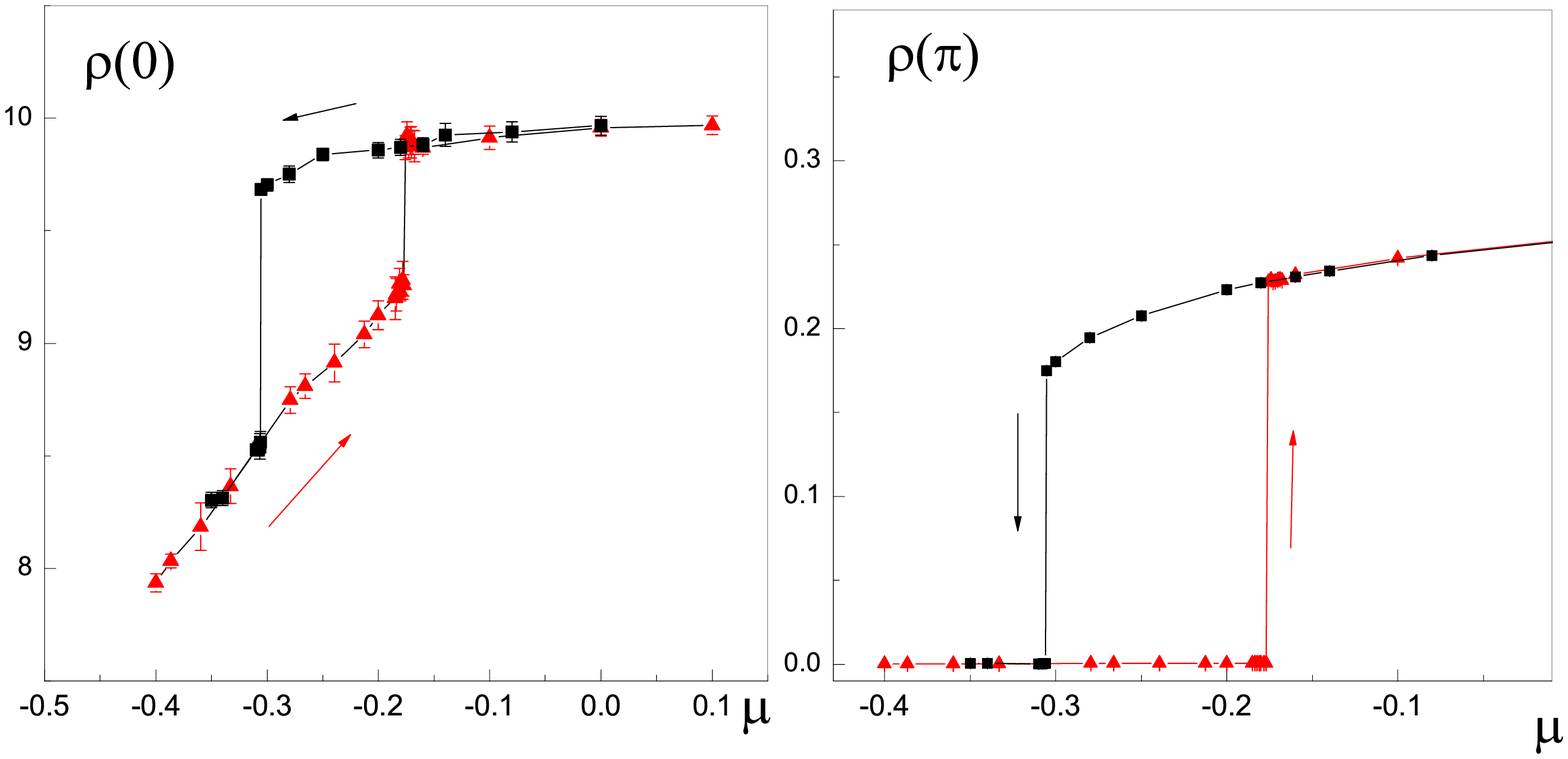}
\end{center}
\vspace{-2.0 cm}
\caption{(Color online)
Hysteresis in the CSF stiffness $ \rho(q_z=0)$ (left pannel) and the counterflow superfluid stiffness $\rho(q_z=\pi)$ vs chemical potenial $\mu$ in the 2d-layers case. 
Simulations have been conducted for the sizes $L=24$, $\beta =24$, $N=8$, and for $K_2=1.65$.
} \label{Iorder}
\end{figure}

We have found the transition to be first-order in 2d and second order in 1d. Larger errorbars in 2d (the upper curve on the main panel, Fig.~\ref{PD}) are due to the hysteretic nature of the transition (shown in Fig.~\ref{Iorder}). Moreover data for $0<n<0.1$ (and $0.9<n<1$) cannot be obtained reliably due to strong finite size effects -- the coherence length $\xi$ at $q_z\neq0 $ diverges exponentially as $K_2 \to 0$ in full analogy to the divergence of the wave function radius of a particle in a bound state of a weak potential in 2d. In other words, the single chain 
size $\sqrt{G(N)}$ becomes exponentially large, so that strong finite size effects are observed
(lines in these regions are guides to eyes indicating that the transition value of $K_2$ must vanish  in the limit of a single chain \cite{Schain}). 
\\ \indent 
We have found that, in contrast with the single chain case, there is
no roughening transition as  $K_2$ increases -- the gyration radius is always diverging with $N$. In other words, chains are always rough. However, such divergence is not characterized by simple $\sim \ln N$ growth at large $N$. Instead, the   $G(N) \sim \ln N$ dependence crosses over to a sub-logarithmical behavior for $N$ larger than some $N^*$. This is shown in the insets of Fig.~\ref{PD}, where $G(N)$ is plotted as a function of the number of layers $N$, in 1d (right) and 2d (left). Fits to numerical data clearly reveal the crossover  at $N_{*} \leq 30-40$: The $\ln N$ behavior (fitted by solid lines) for $N<N_*$ becomes $ G(N) \sim \ln^c (N)$ (dashed lines) for $N>N_*$, with  $c\approx 0.1$ in 1d, and  $c \approx 0.2$ in 2d.   
Correspondingly, the superfluid stiffness exponentially decreases with $N$ as the system crosses over to the sub-log behavior (see Fig.~\ref{rho} below), as if chains were entering some sort of non-superfluid phase due to the {\it caging} effect, i.e., inter-chain repulsion induced by their shape fluctuations \cite{KITPC}. 
\\ \indent
Fig.~\ref{rho} demonstrates how the transition points have been determined (here data refer to 1d case and   $n=0.5$). The main plot shows  $\rho^{(1)}(q_z=\pi)$, as a function of $K_2$, with the transition point $K_2 =(K_2)_{\rm c} \approx 0.42 \pm 0.03$ marked by the vertical arrow. The inset shows the chain superfluid stiffness $\rho^{(1)}(q_z=0)$ for several values of $N$. In the CSF we identify two regimes: $(K_2)_{\rm c} < K_2 \leq 0.6$, and $K_2 \geq 1$. In the latter, the chain binding energy $V_b$ is much larger than the single-molecule tunneling $J$,
and tunneling of a whole chain proceeds as a process of $N$-th order in the ratio $J/V_b <<1$.
This results in an exponential suppression $\rho^{(1)}(q_z=0) \sim \exp(- N/13.2)$. Accordingly, the temperature scale below which $\rho^{(1)}(q_z=0) \neq 0$
is also exponentially suppressed.
In the region $(K_2)_{\rm c} < K_2 \leq 0.6$ the stiffness $ \rho^{(1)}(q_z=0)$ exhibits 
peaks and significant fluctuations characterized by large error bars (for $N>10$ ) and non-exponential dependence on $N$. The exact nature of such fluctuations is an open question. 
\subsection{Detecting chains}
Chains formation is characterized by two qualitative features 
in absorptive imaging: the first is vanishing of the single-particle interference pattern (observed after the release of independent condensates); the second, a radical change of the structure of the density fluctuations. Emergence of these features can be observed as $V_d/J$ is increased above the critical value for a given density $n$. 

The first feature is due to the loss of ODLRO in the single particle density matrix $D_1$.  
The second one is induced by locking of the densities in each layer. Indeed, if no chains are present, the columnar absorptive image 
is characterized by density fluctuations $\Delta n_{tot} \sim \sqrt{Nn}$, where $Nn$ stands for the total number of molecules. When chains are formed, fluctuations occur mostly in their center of mass position, that is, $\Delta n_{tot} \sim \sqrt{n}$ deep in the superfluid chain phase.


\section{Inter-layer non-viscous drag}
It is useful to present an interpretation of the system within the framework of  Andreev-Bashkin effect of non-viscous drag between two condensed components\cite{AB} for the 2d case. Such drag has been shown to lead to exotic vortices in two-component bosonic systems \cite{Kaurov,Egor}. In the present context there are $N$ such components each described by the phase $\varphi_\alpha(\vec{x})$ depending on the space coordinates $\vec{x}$ and the layer index $\alpha=1,2,...N$.    
The corresponding kinetic energy of the superflows can be written as an extension of Eq.(1) from Ref.\cite{Kaurov}: 
\begin{eqnarray}
E_{AB}= \frac{1}{2}\int d^{2} x \sum_{\alpha,\alpha'}  \tilde{\rho}_{\alpha,\alpha'}\vec{\nabla} \varphi_\alpha(\vec{x})   \vec{\nabla} \varphi_{\alpha'}(\vec{x}),
\label{SAB}
\end{eqnarray}
where $\tilde{\rho}_{\alpha,\alpha'}$ are the effective superfluid stiffnesses, (written in a form which takes into account the equivalence of layers as well as their isotropy, i.e. $\rho_{\alpha,\alpha'}$ is a scalar and in general depends on the distance between layers $(b_z|\alpha - \alpha'|)$. In the case of non-interacting layers ($V_d=0$), the N-SF phase is characterized by
$\tilde{\rho}_{\alpha ,\alpha'}=\rho_0 \delta_{\alpha,\alpha'}$ with $\rho_0$ denoting intra-layer stiffnesses (the same on each layer). 
When the dipole interaction is turned on, the
finite inter-layer drag emerges. This results in a non-trivial dependence of  $\tilde{\rho}_{\alpha, \alpha'}$ on $(b_z|\alpha - \alpha'|)$ manifesting a strongly interacting N-SF phase. 

In the CSF phase, the full locking of the superflows from different layers implies that $\tilde{\rho}_{\alpha ,\alpha'}$ becomes
{\it infinitely} ranged along $z$. In other words,  $\tilde{\rho}_{\alpha,\alpha'}$ is independent on the layer indices and is equal to the quantity carrying the meaning of the CSF superfluid stiffness $\rho^{(1)}(0)$ presented in Fig.~\ref{rho}. Resorting to the discrete Fourier transform $\rho(q_z)$ of $\tilde{\rho}(z)$, the CSF phase is characterized by $\rho (q_z)=\rho^{(1)}(0)\delta_{q_z,0}$ (where $q_z=2\pi n_z/N,\, n_z=0,1,2,..., N-1$ for periodic boundary conditions along $z$), that is the N-SF to SCF transition is characterized by emergence of {\it infinitely}-ranged non-viscous drag between layers.

\subsection{The mean field approximation}
Within the mean field description, the relevant interaction term responsible for the discussed quantum phase transition is $ \sim \langle \Phi^*\rangle \langle \psi \rangle^N + c.c. $,  where $ \langle \Phi \rangle$ stands for the chain order parameter and  
$\langle \psi \rangle$ describes molecular condensates in each layer. One can therefore write down the minimal Landau free energy in terms of $\psi$ and $\Phi$ fields:
\begin{equation}
{\it F}=a_1|\psi|^2+a_2|\Phi|^2+b_1|\psi|^4+b_2|\Phi|^4- c_{12} (\Phi^*\psi^N+{\rm cc}),
\label{free_E}
\end{equation}
where $a_{1,2}$ , $b_{1,2},\,c_{12}$ are some phenomenological coefficients. For $N>2$ this functional has relevant negative term of order higher than 4, 
implying discontinuous transition between the N-SF state, where $\psi \neq 0$ (and trivially $\Phi \neq 0$), and the CSF where $\psi=0$ and $\Phi \neq 0$.
This is consistent with our observation of strong first-order quantum phase transition in 2d (at $T=0$), 
with the typical hysteretic behavior demonstrated in Fig.~\ref{Iorder}. 
It is interesting to investigate how the first-order transition transforms at finite $T$ in 2d when the vortex proliferation is characterized by strong drag effect
as decsribed above.   

In D=1+1 (at $T=0$), the mean field description is not applicable. Instead, the bosonization approach \cite{Ashvin,Mathey,KITPC} describing Berezinskii-Kosterlitz-Thouless scenario is more appropriate. Accordingly, no hysteresis has been observed in our simulations in the case of 1d "cigars".

\section{Discussion and summary}

Polarized bosonic molecules confined in a stack of $N$ identical one- and two- dimensional optical lattice layers form 
a chain superfluid phase -- CSF at large enough dipolar interaction. Such phase is characterized by ODLRO only in the $N$-body density matrix. 
Moreover, it is proved that no superfluidity of chains shorter than $N$ can exist, provided all layers are identical to each other.

Upon decreasing the interaction strength the system undergoes a quantum phase transition from CSF to the $N$-independent molecular superfluids. In 2d this transition is of strongly first-order (for $N>2$), while in the
1d case, it is continuous and can be interpreted as a multi-layer version of the Berezinskii-Kosterlitz-Thouless transition. 

In terms of an effective description, the quantum phase transition in the 2d case is characterized by emergence of the non-viscous inter-layer drag effect
with infinite range correlations between layers. An interesting question is about the nature of vortices and their unbinding in the N-SF phase close to the CSF at finite $T$. Similarly to the 2-component case \cite{Kaurov,Egor}, such vortices should be composite objects. A closely related question concerns the fate of the first-order transition at finite $T$.    

We have also studied single chain behavior by means of quantum Monte Carlo simulations, and have observed the quantum roughening transition from a regime of stiff chain, at large dipolar interaction strength, to rough chain, at smaller strengths. At zero temperature the rough chain behavior is characterized by logarithmical divergence of the gyration radius with the number of layers $N$. In the stiff (smooth) phase the gyration radius becomes independent of $N$. 

While no roughening transition was found in the CSF, the single-chain transition leaves its "footprint" in the CSF phase: the divergence of the gyration radius as a function of $ N$ crosses over from  $\sim \ln N$ to  $\sim \ln^c N$ type, where $c\approx 0.1-0.2$ (in contrast with the value $c=0$ for a single chain). In such a regime chains stay rough, which means that CSF is always a superfluid of {\it flexible} chains). Crystalline order of chains, which would imply loss of roughness, though, is expected to appear for strong enough dipolar interaction once intra-layer repulsion is considered.

In future we would like to investigate the interplay between rough chain superfluidity, independent molecular superfluids and solid phases by describing the system using the full quantum hamiltonian. In particular,  the possibility of a supersolid phase in such systems will be explored. Another interesting question concerns how finite tunneling between layers (which has been neglected in the present analysis) can affect the system.

As far as the practical realization of the chain superfluid phase is concerned, we note that our analysis valid for bosonic molecules can be applied to the case of fermionic molecules (KRb \cite{Jila1,Jila2}) as well with some modifications. Specifically, the symmetry between layers should be broken by grouping them into pairs, so that
the distance between layers within each pair is smaller than the distance between the pairs. In this case, it should be possible to achieve strong pairing of fermionic
molecules from each pair of layers into essentially bosonic di-molecules, which in their turn can form chains.   

\begin{acknowledgements}
Authors are grateful to  Yu.E. Lozovik, N.V. Prokof'ev, B.V. Svistunov and M. Troyer  
for stimulating discussions and to D. Jin for discussing experimental 
aspects. One of us (A.B.K.) is thankfull to KITPC at Beijing, KITP at Santa Barbara and Nordita for hospitality.
This work was supported by the Institute for Theoretical Atomic, Molecular and Optical Physics (ITAMP), and the National Science Foundation
under Grant No.PHY1005527, CUNY PSC grant 63071-0041 and by a grant of computer time from the CUNY HPCC under NSF Grants CNS-0855217 and CNS - 0958379.
\end{acknowledgements}


\end{document}